\documentclass[twocolumn,showpacs,preprintnumbers,amsmath,amssymb]{revtex4}

\usepackage{hyperref}
\usepackage{graphicx}
\usepackage{dcolumn}
\usepackage{bm}

\begin{document}

\title{Flat manifold leptogenesis in the supersymmetric standard model}

\author{Masato Senami}
\altaffiliation{E-mail address: senami@nucleng.kyoto-u.ac.jp}
\author{Katsuji Yamamoto}
\altaffiliation{E-mail address: yamamoto@nucleng.kyoto-u.ac.jp}
 \affiliation{Department of Nuclear Engineering, Kyoto University,
Kyoto 606-8501, Japan}

\date{\today}

\begin{abstract}
Flat manifold leptogenesis a la Affleck-Dine
is investigated with the slepton and Higgs fields,
$ \tilde{L} $, $ H_u $, $ H_d $,
in the supersymmetric standard model.
The multi-dimensional motion of these scalar fields
is realized in the case that the $ \tilde{L} H_u $ and $ H_u H_d $ directions
are comparably flat with the relevant non-renormalizable
superpotential terms.
Soon after the inflation, the lepton number asymmetry
appears to fluctuate due to this multi-dimensional motion
involving certain $ CP $ violating phases.
Then, it is fixed to some significant non-zero value
for the successful baryogenesis
when the scalar fields begin to oscillate with rotating phases
driven by the quartic coupling from the superpotential term
$ {\bar h}_e L H_d e^c $ with $ {\bar h}_e \sim 10^{-5} - 10^{-3} $.
The Hubble parameter  $ H_{\rm osc} $ at this epoch
for the completion of leptogenesis is much larger
than the gravitino mass $ m_{3/2} \sim 10^3 {\rm GeV} $.
The thermal terms may even play a cooperative role
in this scenario of early leptogenesis.
The lightest neutrino mass can be $ m_{\nu_1} \sim 10^{-4} {\rm eV} $,
if the reheating temperature is allowed to be
$ T_R \sim 10^{10} {\rm GeV} $.
\end{abstract}
\maketitle

\section{Introduction}
\label{sec:introduction}

The Affleck-Dine mechanism \cite{AD,DRT} is well known
as one of the promising candidates
to explain the long-standing cosmological problem, baryogenesis,
in supersymmetric models.
Particularly, the $ \tilde{L} H_u$ flat direction
has been investigated extensively for leptogenesis
in the supersymmetric standard model,
where the generated lepton number asymmetry is converted
with a significant fraction to the baryon number asymmetry
through the anomalous electroweak process
\cite{DRT,LHu,LHu2,LHu3}.
In this $ \tilde{L} H_u $ case, there appears an interesting relation
between the baryon number asymmetry and the neutrino masses
through the superpotential term $ ( L H_u )( L H_u ) $.
It should here be noticed that the $ ( H_u H_d )( H_u H_d ) $ term
is implicitly assumed to be larger than the $ ( L H_u )( L H_u ) $ term.
Otherwise, the $ H_u H_d $ would be the flattest direction
without producing the lepton number asymmetry.

In this article, we reexamine the leptogenesis
a la Affleck-Dine in the supersymmetric standard model
by considering the new possibility that a flat manifold is formed
for the $ \tilde{L} $, $ H_u $ and $ H_d $ fields
with comparable $ ( L H_u )( L H_u ) $ and $ ( H_u H_d )( H_u H_d ) $ terms.
This comparability seems to be plausible if these non-renormalizable terms
stem from the physics at the Planck scale.
It has been observed recently
in an extension of the supersymmetric standard model with triplet Higgs
that the multi-dimensional motion can really be realized
for the scalar fields on a flat manifold
\cite{myletter}.
In the multi-dimensional motion on the flat manifold,
non-conservation of certain particle numbers
with effective $ CP $ violation
may be available from some potential terms
with different dependences on the scalar field phases.
Then, soon after the inflation
the fluctuating motion of the scalar fields appears
due to the effects of these potential terms,
and the particle number asymmetries
such as the $ B - L $ asymmetry are generated varying in time.
We will show in the text that this sort of phenomenon indeed occurs
for the flat manifold leptogenesis
with the scalar fields $ \tilde{L} $, $ H_u $ and $ H_d $
in the supersymmetric standard model.
It is the novel point in the present scenario
that the lepton number asymmetry fluctuating after the inflation
is fixed to some significant non-zero value
due to the effect of the superpotential term $ {\bar h}_e L H_d e^c $.
While the quartic term $ {\bar h}_e^2 | {\tilde L} H_d |^2 $
is safely small during the inflation
with $ {\bar h}_e \sim 10^{-5} - 10^{-3} $,
it in turn provides the driving force
for the scalar fields to oscillate with rotating phases
at some epoch after the inflation.
The Hubble parameter $ H_{\rm osc} $ at this epoch
of the onset of oscillation by the $ {\bar h}_e $ quartic term
can be much larger than the gravitino mass $ m_{3/2} \sim 10^3 {\rm GeV} $.
The leptogenesis is completed in this quite early epoch
with $ H \sim H_{\rm osc} $, which may even be
before the thermal terms \cite{thermaleffect,LHu2,LHu3}
become significant.
Hence, this flat manifold leptogenesis is not restricted
by the physics at the electroweak scale
such as the low-energy supersymmetry breaking terms.
This is in salient contrast to the conventional flat direction leptogenesis.

This article is organized as follows.
In Sec. \ref{sec:model}, the relevant part
of the supersymmetric standard model is presented
for the flat manifold leptogenesis.
In Sec. \ref{sec:leptogenesis}, the mechanism of flat manifold leptogenesis
is described following the respective epochs starting with the inflation,
and the resultant lepton number asymmetry is estimated
specifically related to the lightest neutrino mass.
In Sec. \ref{sec:thermal}, the thermal effects are discussed
to show that they do not alter essentially
the present scenario for leptogenesis.
In Sec. \ref{sec:numerical}, detailed numerical calculations are made
to confirm the multi-dimensional motion of the scalar fields,
and the reasonable parameter range
for the sufficient leptogenesis is identified.
Sec. \ref{sec:conclusion} is finally devoted to the conclusion
of this investigation for the flat manifold leptogenesis.

\section{Model}
\label{sec:model}

We investigate the supersymmetric standard model
including the non-renormalizable superpotential terms,
\begin{eqnarray}
W_{\rm non} = \frac{\lambda_{{L \! \! \! /}_i}}{2M} ( L_i H_u )( L_i H_u )
            + \frac{\lambda_H}{2M} ( H_u H_d )( H_u H_d ) ,
\label{Wnon}
\end{eqnarray}
where $ M $ represents some very large mass scale such as the Planck scale,
and the suitable basis is chosen for the relevant fields
to give the positive and diagonal $ \lambda_{{L \! \! \! /}_i} $
and positive $ \lambda_H $.
The $ R $-parity violating terms $ ( L_i H_u )( H_u H_d ) $
are not included for simplicity.
These terms do not alter the present leptogenesis scenario
if they are not significantly large compared to $ W_{\rm non} $.
We here assume the condition on the terms in $ W_{\rm non} $,
\begin{equation}
\lambda_H \sim \lambda_{{L \! \! \! /}_1}
\ll \lambda_{{L \! \! \! /}_2} , \lambda_{{L \! \! \! /}_3} .
\label{lamcond}
\end{equation}
Then, the flat manifold for leptogenesis is formed
with the $ {\tilde L}_1 $, $ H_u $ and $ H_d $ fields.
The lepton doublet $ L_1 $ includes the lightest neutrino $ \nu_1 $ with mass
\begin{eqnarray}
m_{\nu_1} = \lambda_{{L \! \! \! /}_1} \frac{\langle H_u \rangle^2}{M}
\sim 10^{-6} {\rm eV}
\left( \frac{10^{19} {\rm GeV}}{M / \lambda_{{L \! \! \! /}_1}} \right) ,
\label{nmass}
\end{eqnarray}
where $ {\sqrt{\langle H_u \rangle^2 + \langle H_d \rangle^2}}
= 174 {\rm GeV} $
with $ \langle H_u \rangle / \langle H_d \rangle \equiv \tan \beta $.
The comparability of $ \lambda_{{L \! \! \! /}_1} $ and $ \lambda_H $
would be understood by considering that these non-renormalizable terms
stem from the physics at the Planck scale.
On the other hand, as indicated from the solar and atmospheric neutrino
puzzles, the other two neutrino masses are probably
around $ 10^{-2} {\rm eV} $, apparently requiring
$ \lambda_{{L \! \! \! /}_2} , \lambda_{{L \! \! \! /}_3}
\sim 10^4 ( M / 10^{19} {\rm GeV} ) $.
Such large couplings may be obtained effectively
as $ \lambda_{{L \! \! \! /}_2} , \lambda_{{L \! \! \! /}_3}
\sim \lambda_{{L \! \! \! /}_1} ( M /  M_{\nu^c} ) $
by the so-called seesaw mechanism with heavy right-handed neutrinos
at an intermediate scale $ M_{\nu^c} \ll M $
\cite{seesaw}.
The see-saw contribution to $ \lambda_{{L \! \! \! /}_1} / M $ should,
however, be suppressed sufficiently for the successful leptogenesis.

The superpotential terms relevant for the flat manifold leptogenesis
are given by
\begin{equation}
W = ( h_e )_{ij} L_i H_d e_j^c + \mu H_u H_d + W_{\rm non} .
\label{W}
\end{equation}
The $ F $ terms except for the contributions of $ W_{\rm non} $
are then calculated as
\begin{eqnarray}
&& F_{L_i} = ( h_e )_{ij} H_d e_j^c , F_{e_j^c} = ( h_e )_{ij} L_i H_d ,
\nonumber \\
&& F_{H_u} = \mu H_d , F_{H_d} = ( h_e )_{ij} L_i e_j^c + \mu H_u .
\end{eqnarray}
Among the slepton fields, only the sneutrino $ {\tilde \nu}_1 $
in $ {\tilde L}_1 $ associated with the lightest neutrino
is considered to develop a very large coherent field value
during the inflation, according to the condition (\ref{lamcond}).
Then, it is usually supposed that the $ {\tilde \nu}_1 $ and $ H_d^0 $
are incompatible for the flat directions due to the $ F_{e^c} $ terms.
However, this is not necessarily the case in the present scenario.
In fact, the $ F_{e^c} $ terms provide the quartic term
\begin{equation}
{\bar h}_e^2 | {\tilde L}_1 H_d |^2
\end{equation}
with
\begin{equation}
{\bar h}_e^2 \equiv \sum_j | ( h_e )_{1j} |^2 .
\end{equation}
The effective coupling is estimated as
\begin{equation}
{\bar h}_e \simeq \frac{m_e + \sin \theta_{12} m_\mu}{\langle H_d \rangle}
\sim \left\{ \begin{array}{ll}
10^{-5} & ( \theta_{12} \lesssim 10^{-2} ) \\
10^{-3} & ( \theta_{12} \sim 1 )
\end{array} \right. ,
\label{h_e}
\end{equation}
where $ \theta_{12} $ represents the $ \nu_e $-$ \nu_\mu $ mixing,
and the effect of the $ \nu_e $-$ \nu_\tau $ mixing
is assumed to be small enough.
As explained in Sec. \ref{sec:leptogenesis},
while this quartic term is safely small during the inflation
with $ {\bar h}_e \sim 10^{-5} - 10^{-3} $,
it in turn provides the driving force
for the scalar fields to oscillate with rotating phases
at some epoch after the inflation.
This is one of the essential points in the present scenario
for the flat manifold leptogenesis.

According to the above arguments, the flat manifold is specified
by the $ D $-flat condition
for the $ {\rm SU(2)}_{\rm L} \times {\rm U(1)}_Y $ gauge interactions,
\begin{equation}
| {\tilde \nu}_1 |^2 - | H_u^0 |^2 + | H_d^0 |^2 = 0 ,
\label{Dflat}
\end{equation}
and the other fields are vanishing.
Henceforth, we adopt for simplicity the notation,
$ {\tilde L} \equiv {\tilde \nu}_1 $,
$ H_u \equiv H_u^0 $ and $ H_d \equiv H_d^0 $,
suppressing the lepton generation indices.
Including the contributions of $ W_{\rm non} $,
this manifold is flat enough for both the $ {\tilde L} H_u $ and $ H_u H_d $
directions in the case that $ \lambda_{L \! \! \! /} $ and $ \lambda_H $
are comparable,
\begin{equation}
0.3 \lesssim \lambda_{L \! \! \! /} /\lambda_H \lesssim 3 ,
\label{flatcond}
\end{equation}
as observed in Ref. \cite{myletter}.

\section{Flat Manifold Leptogenesis}
\label{sec:leptogenesis}

We here describe the mechanism of leptogenesis
on the flat manifold consisting of the scalar fields
$ {\tilde L} , H_u , H_d $, say {\it AD-flatons}.
(The scalar fields associated with the flat potential
are intrinsic in supersymmetric models,
which are named flatons \cite{flaton}.
Here, we consider such fields as participants
in leptogenesis/baryogenesis a la Affleck-Dine.)
The scalar potential for the AD-flatons
$ \phi_a = {\tilde L} , H_u , H_d $ may be presented as
\begin{equation}
V = V_{\rm high} + {\bar h}_e^2 | {\tilde L} H_d |^2 + V_{\rm low} .
\label{V}
\end{equation}
These three parts become dominant in the high, middle
and low energy scales, respectively.
The evolution of the AD-flaton fields
is traced in the corresponding epochs starting with the inflation.
The high-energy part is provided from the superpotential $ W_{\rm non} $
with the mass scale $ M $ and the corresponding soft supersymmetry breaking
with the Hubble parameter $ H $:
\begin{eqnarray}
V_{\rm high} &=& - \sum_a c_a H^2 | \phi_a |^2
\nonumber \\
&+& \left| \lambda_{L \! \! \! /} \frac{\tilde{L} ( \tilde{L} H_u )}{M}
 + \lambda_H \frac{H_d ( H_u  H_d )}{M}  \right|^2
\nonumber \\
&+& \left| \lambda_H \frac{H_u ( H_u H_d )}{M} \right|^2
 + \left| \lambda_{L \! \! \! /} \frac{ H_u ( \tilde{L} H_u )}{M}\right|^2
\nonumber \\
&+& \left[ \frac{\lambda_{L \! \! \! /}}{2M} a_{L \! \! \! /} H
           ( \tilde{L} H_u )( \tilde{L} H_u ) + {\rm h.c.} \right]
\nonumber \\
&+& \left[ \frac{ \lambda_H}{2M} a_H H
           ( H_u H_d )( H_u H_d ) + {\rm h.c.} \right]
\nonumber \\
&+& ( g^2 + g^{\prime 2} ) ( | {\tilde L} |^2 - | H_u |^2 + | H_d |^2 )^2 .
\label{Vhigh}
\end{eqnarray}
The last term with the gauge couplings $ g $ and $ g^\prime $
of the $ {\rm SU(2)}_{\rm L} $ and $ {\rm U(1)}_Y $, respectively,
is also included here to realize the $ D $-flat condition (\ref{Dflat})
for the very large $ | {\tilde L} | $, $ | H_u | $ and $ | H_d | $.
The low-energy part $ V_{\rm low} $ includes the remaining terms,
which are related to the parameters
$ m_{3/2} , | \mu | \sim 10^3 {\rm GeV} $.
It will be clarified in the following
that the quartic term $ {\bar h}_e^2 | {\tilde L} H_d |^2 $
plays a crucial role to complete the flat manifold leptogenesis.
On the other hand, the low-energy part $ V_{\rm low} $
is actually irrelevant for the leptogenesis.
This is in salient contrast to the conventional flat direction
Affleck-Dine mechanism.
In this section, we do not either consider explicitly
the thermal terms \cite{thermaleffect,LHu2,LHu3}.
It will be seen in Secs. \ref{sec:thermal} and \ref{sec:numerical}
that the thermal terms do not alter essentially
the present scenario of flat manifold leptogenesis.
They may even play a cooperative role for leptogenesis.

\subsection{$ H = H_{\rm inf} $}

During the inflation the Hubble parameter takes almost a constant value
$ H_{\rm inf} $, which is typically $ 10^{14} {\rm GeV} $ or so.
In this epoch, the AD-flaton fields quickly settle into one of the minima
of the scalar potential (\ref{V}) with $ H = H_{\rm inf} $.
The potential minima is almost determined
by the high-energy part $ V_{\rm high} $ as
\begin{equation}
\phi_a^{(0)} = {\rm e}^{i \theta_a^{(0)}} r_a^{(0)}
\sqrt{{H_{\rm inf} (M/\lambda)}} ,
\label{phi0}
\end{equation}
with $ \lambda $ representing the mean value
of $ \lambda_{L \! \! \! /} $ and $ \lambda_H $.
The generation of these non-trivial minima (\ref{phi0})
with $ r_a^{(0)} \sim 0.1 - 1 $ far apart from the origin
on the multi-dimensional flat manifold depends rather complicatedly
on the parameters, $ c_a $, $ \lambda_{L \! \! \! /} $, $ \lambda_H $,
$ a_{L \! \! \! /} $ and $ a_H $
\cite{myletter}.
As usually considered, at least one of $ c_a $'s should be positive
so that the origin stays unstable in the inflation epoch.
It is also essential for the flat manifold formation
that $ \lambda_{L \! \! \! /} $ and $ \lambda_H $ are comparable,
as given in Eq. (\ref{flatcond}).
Otherwise, the potential minima would be developed
along either the $ {\tilde L} H_u $ direction or the $ H_u H_d $ direction.

By comparing $ V_{\rm high} \sim H_{\rm inf}^3 ( M / \lambda ) $ and
$ {\bar h}_e^2 | {\tilde L} H_d |^2
\sim {\bar h}_e^2 H_{\rm inf}^2 ( M / \lambda )^2 $,
the following condition is expected to be satisfied
for the flatness of the $ {\tilde L} $-$ H_u $-$ H_d $ manifold
during the inflation:
\begin{equation}
H_{\rm inf} > H_{\rm osc} \equiv {\bar h}_e^2 ( M / \lambda ) .
\label{Hinfcond}
\end{equation}
The critical value $ H_{\rm osc} $ of the Hubble parameter
for which the $ {\bar h}_e $ quartic term becomes comparable
to $ V_{\rm high} $ may be estimated in terms of $ m_{\nu_1} $
with Eqs. (\ref{nmass}) and (\ref{h_e}):
\begin{equation}
H_{\rm osc} \sim 10^9 {\rm GeV}
\left( \frac{{\bar h}_e}{10^{-5}} \right)^2
\left( \frac{10^{-6} \rm eV}{m_{\nu_1}} \right) .
\label{Hosc}
\end{equation}
Hence, the condition (\ref{Hinfcond}) can be satisfied readily
since $ H_{\rm inf} $ is typically $ 10^{14} {\rm GeV} $ or so.
Here, it may be noticed that if $ {\bar h}_e \sim 10^{-3} $
with large $ \nu_e $-$ \nu_\mu $ mixing,
$ H_{\rm osc} $ could be comparable to $ H_{\rm inf} $.
Remarkably, even in such an extreme case
with $ H_{\rm osc} \sim H_{\rm inf} $
the successful leptogenesis can be realized,
as confirmed by the numerical calculations in Sec. \ref{sec:numerical}.

\subsection{$ H_{\rm osc} < H < H_{\rm inf}  $}

After the inflation the inflaton oscillates coherently,
and it dominates the energy density of the universe.
In this epoch with $ H > H_{\rm osc} $,
the high-energy part $ V_{\rm high} $ is still dominant,
and the evolution of the AD-flaton fields is essentially
the same as in the case of the model with triplet Higgs
\cite{myletter}.
We recapitulate the main results with suitable changes of notation,
showing especially the fluctuating behavior of the lepton number asymmetry
after the inflation.

The AD-flaton fields are moving toward the origin
with the initial conditions at $ t = t_0 \sim H_{\rm inf}^{-1}$
after the inflation,
\begin{equation}
\phi_a (t_0) = \phi_a^{(0)} , \ {\dot \phi}_a (t_0) = 0 .
\label{phit0}
\end{equation}
The evolution of the AD-flaton fields is governed
by the equations of motion,
\begin{equation}
\ddot{\phi}_a + 3 H \dot{\phi}_a
+ \frac{\partial V}{\partial \phi_a^*} = 0 .
\label{eqofm}
\end{equation}
The Hubble parameter varies in time as $ H = (2/3) t^{-1} $
in the matter-dominated universe.
The AD-flaton fields may be represented suitably
in terms of the dimensionless fields $ \chi_a $
\cite{DRT} as
\begin{eqnarray}
\phi_a = \chi_a \sqrt{{H (M/\lambda)}}
\equiv {\rm e}^{i \theta_a} r_a \sqrt{{H (M/\lambda)}} .
\label{phi}
\end{eqnarray}
Then, the equations of motion (\ref {eqofm}) are rewritten
with $ z = \ln ( t / t_0 ) $ as
\begin{equation}
\frac{d^2 \chi_a}{dz^2} + \frac{\partial U}{\partial \chi_a^*} = 0 ,
\label{eqofmchi}
\end{equation}
and the initial conditions from Eq. (\ref{phit0}) are given as
\begin{equation}
\chi_a (0) = {\rm e}^{i \theta_a^{(0)}} r_a^{(0)} , \
\frac{d \chi_a}{dz} (0) = \frac{1}{2} \chi_a (0) .
\label{chi0}
\end{equation}
It should be noticed in Eq. (\ref{eqofmchi})
that the first-order $ z $-derivative is absent
due to the parametrization of $ \phi_a $ in Eq. (\ref{phi}).
The dimensionless effective potential is given dominantly by
\begin{eqnarray}
U ( \chi_a )
& \simeq & \frac{4}{9 H^3 ( M / \lambda)} V_{\rm high}
- \frac{1}{4} | \chi_a |^2
\nonumber \\
&+& \frac{4}{9} ( H_{\rm osc} / H ) | \chi_{\tilde L} \chi_{H_d} |^2 .
\label{U}
\end{eqnarray}
The second term is due to the time variation of the factor
$ \sqrt{{H (M/\lambda)}} $ in Eq. (\ref{phi}),
which apparently provides the change of the mass terms in $ U ( \chi_a ) $,
\begin{equation}
c_a \rightarrow c_a + \frac{9}{16} .
\label{dca}
\end{equation}
The third contribution from the $ {\bar h}_e $ quartic term
is small enough in this epoch with $ H > H_{\rm osc} $,
while it will in turn play a crucial role in the next epoch.
The low-energy part $ V_{\rm low} $ is clearly negligible
for $ H_{\rm osc} \gg m_{3/2} , | \mu | $.

The behavior of the AD-flaton phases $ \theta_a $
is described in this epoch as follows. 
The initial conditions at $ t = t_0 $ ($ z = 0 $)
are given from Eq. (\ref{chi0}) as
\begin{equation}
\theta_a (0) = \theta_a^{(0)} , \ \frac{d \theta_a}{dz} (0) = 0 .
\end{equation}
On the other hand, the asymptotic trajectory of the AD-flaton fields
is found by the conditions $ \partial U / \partial \chi_a^* = 0 $
in this epoch with the small enough $ {\bar h}_e $ quartic term as
\begin{equation}
\theta_a = \theta_a^{(1)} .
\end{equation}
It is remarkable in the multi-dimensional motion
of the AD-flaton fields with $ \lambda_{L \! \! \! /} \sim \lambda_H $
that the direction of this trajectory is somewhat different
from the initial direction, i.e.,
\begin{equation}
\theta_a^{(1)} \not= \theta_a^{(0)} .
\end{equation}
This is because the apparent change of the mass terms in Eq. (\ref{dca})
due to the redshift induces the new balance
among the $ \lambda_{L \! \! \! /} $-$ \lambda_H $ cross term,
$ a_{L \! \! \! /} $ term and $ a_H $ term in $ U ( \chi_a ) $,
which have different dependences on $ \theta_a $.
(If the fine-tuning is made as
$ \arg ( a_{L \! \! \! /} ) - \arg ( a_H ) = \pi \ {\rm mod} \ 2 \pi $,
the initial balance is maintained independently of $ | \chi_a | $
so as to realize $ \theta_a^{(0)} = \theta_a^{(1)} $.)
Without the $ d \chi_a / d z $ (friction) term in Eq. (\ref{eqofmchi}),
the AD-flaton phases $ \theta_a $ slowly fluctuate
around $ \theta_a^{(1)} $ starting from $ \theta_a^{(0)} $
as a function of $ z = \ln ( t / t_0 ) $
in the epoch $ H_{\rm inf}^{-1} \sim t_0 \leq t < H_{\rm osc}^{-1} $.
That is, in the motion on the multi-dimensional flat manifold

the AD-flaton fields no longer track exactly behind the decreasing
instantaneous minimum of the scalar potential.
This is in salient contrast to the conventional Affleck-Dine mechanism
on the one-dimensional flat direction,
where the AD-flaton phase is fixed for a long time
until the low-energy supersymmetry breaking terms become important.

In this way, through this fluctuating motion
of the (almost) homogeneous coherent AD-flaton fields $ \phi_a (t) $,
the particle number asymmetries are generated
soon after the inflation as
\begin{eqnarray}
\Delta n_a \equiv n_a - {\bar n}_a
= i ( \phi_a^* {\dot \phi}_a - {\dot \phi}_a^* \phi_a ) .
\end{eqnarray}
The fractions of the respective asymmetries are also calculated
by considering the redshift in the matter-dominated universe as
\begin{equation}
\epsilon_a (t) \equiv \Delta n_a / [ (3/2) H^2 ( M / \lambda ) ]
= - 2 r_a^2 \frac{d \theta_a}{dz} .
\label{ea}
\end{equation}
The lepton number asymmetry is particularly given
as the $ {\tilde L} $ asymmetry,
\begin{equation}
\epsilon_L (t) = \epsilon_{\tilde L} (t) .
\label{eL}
\end{equation}
Since the AD-flaton phases are fluctuating in this early epoch,
as mentioned so far, the lepton number asymmetry
is oscillating in time as $ | \epsilon_L (t) | \sim | d \theta_a / dz |
\lesssim | \theta_a^{(0)} - \theta_a^{(1)} | \sim 0.01 - 0.1 $
($ r_a \sim 0.1 - 1 $) numerically for the reasonable parameter values.

\subsection{$ m_{3/2} \ll H \lesssim H_{\rm osc} $}

The high-energy potential terms and the $ {\bar h}_e $ quartic term
are redshifted for $ H \gtrsim H_{\rm osc} $ after the inflation as
\begin{eqnarray}
V_{\rm high} & \sim & H^3 ( M / \lambda ) ,
\\
{\bar h}_e^2 | {\tilde L} H_d |^2
& \sim & {\bar h}_e^2 H^2 ( M / \lambda )^2
\end{eqnarray}
with $ | \phi_a | \sim {\sqrt{H ( M / \lambda )}} $ in Eq. (\ref{phi}).
Then, the $ {\bar h}_e $ quartic term eventually dominates
in the present epoch with $ H \lesssim H_{\rm osc} $,
playing the crucial role for the flat manifold leptogenesis.
Specifically, this quartic term acts in some sense
as positive mass-squared terms for the $ {\tilde L } $ and $ H_d $ fields,
driving the AD-flaton fields to oscillate.
Once the oscillation begins,
the lepton number asymmetry is fixed to some non-zero value as
\begin{equation}
\epsilon_L (t) = \epsilon_L \ ( t \gg H_{\rm osc}^{-1} ) .
\label{eL1}
\end{equation}
This asymptotic behavior of $ \epsilon_L (t) $ in the later time
is in accordance with the fact
that the lepton number violating terms in $ V_{\rm high} $
are redshifted rapidly to be much smaller
than the lepton number conserving terms
including the $ {\bar h}_e $ quartic term.
(See Sec. \ref{sec:numerical} for the numerical results.)
It is also obvious that for $ H \gg m_{3/2} , | \mu | $
the low-energy part $ V_{\rm low} $ still provides negligible effects
in this epoch of leptogenesis.

The coherent oscillation of the inflaton field dominates
the energy density of the universe until the decay of inflatons
is completed at the time $ t_R $ ($ \gg H_{\rm osc}^{-1} $).
Then, the universe is reheated to the temperature $ T_R $.
Until this time the lepton number asymmetry is redshifted as matter,
which is given at $ t = t_R $ ($ H = H_R $)
with Eqs. (\ref{ea}), (\ref{eL}) and (\ref{eL1}) as
\begin{eqnarray}
n_L(t_R) = \epsilon_L (3/2) H_R^2 (M/\lambda) .
\end{eqnarray}
Then, the lepton-to-entropy ratio after the reheating
is determined with $ s \simeq 3 H_R^2 M_{\rm P}^2 / T_R $ as
\begin{eqnarray}
\frac{n_L}{s}
& \simeq & \epsilon_L \frac{( M / \lambda ) T_R}{2 M_{\rm P}^2}
\nonumber \\
& \sim & 10^{-10}
    \left( \frac{\epsilon_L}{0.1} \right)
    \left( \frac{10^{-6} {\rm eV}}{m_{\nu_1}} \right)
    \left( \frac{T_R}{10^9 \rm GeV} \right) ,
\label{kinji}
\end{eqnarray}
where $ M_{\rm P} = m_{\rm P} / {\sqrt{8 \pi}}
= 2.4 \times 10^{18} {\rm GeV} $ is the reduced Planck mass,
and the relation (\ref{nmass}) between $ M / \lambda $ and $ m_{\nu_1} $
is considered.
This lepton number asymmetry is converted partially
to the baryon number asymmetry through the electroweak anomalous effect.
The chemical equilibrium between leptons and baryons leads
the ratio $ n_B = - (8/23) n_L $
(without any preexisting baryon number asymmetry)
\cite{harveyturner}.
Therefore, the sufficient baryon-to-entropy ratio can be provided
as required from the nucleosynthesis
with $ \eta = ( 2.6 - 6.2 ) \times 10^{-10} $ \cite{eta}.

It should here be noted that
the reheating temperature may be constrained
as $ T_R \lesssim 10^8 - 10^{10} {\rm GeV} $, or even more severely,
for $ m_{3/2} \sim 1 {\rm TeV} $ to avoid the gravitino problem
\cite{gravitino,gravitino2,gravitino3}.
Hence, the desired mass of the lightest neutrino is very small
generally as $ m_{\nu_1} \lesssim 10^{-6} {\rm eV} $.
It will, however, be found in Sec. \ref{sec:numerical}
that there is some range of the model parameters
for the successful leptogenesis with relatively large
$ m_{\nu_1} \sim 10^{-4} {\rm eV} $,
if the reheating temperature is allowed
to be $ T_R \sim 10^{10} {\rm GeV} $
for $ m_{3/2} \sim {\mbox{several TeV}} $
with small hadronic branching ratio of gravitino decay $ B_h \sim 0.01 $
\cite{gravitino2}.
Then, the prediction for the neutrinoless double beta decay
with very small $ m_{\nu_1} $ \cite{LHu3} could be evaded
in the case of $ {\tilde L} $-$ H_u $-$ H_d $ flat manifold leptogenesis.

\section{Thermal Effects}
\label{sec:thermal}

We now discuss that the thermal terms for the scalar potential
\cite{thermaleffect,LHu2,LHu3} do not alter essentially
the present scenario of flat manifold leptogenesis.
Before the reheating after inflation is completed,
there is already a dilute plasma of the inflaton decay products
with temperature
\begin{equation}
T_{\rm p} \sim ( T_R^2 H M_{\rm P} )^{1/4} .
\end{equation}
Then, the AD-flatons acquire the thermal mass terms in this plasma,
\begin{equation}
V_{\rm th1} = c_{\rm th} y^2 T_{\rm p}^2 | \phi_a |^2 \
( y | \phi_a | < T_{\rm p} ) ,
\end{equation}
where $ y $ is the relevant coupling constant,
and $ c_{\rm th} $ is the positive constant,
e.g., $ c_{\rm th} = 3/4 $ for a quark superfield.
One can readily estimate the Hubble parameter $ H_{\rm th1} $
when the thermal mass terms begin to dominate
over the Hubble induced mass terms:
\begin{equation}
H_{\rm th1} \sim \min \left[ \frac{T_R^2 M_{\rm P}}{y^4 (M / \lambda)^2},
( y^4 T_R^2 M_{\rm P} )^{1/3} \right] .
\end{equation}
The thermal log terms are also given mainly through the modification
of the $ {\rm SU(3)}_C $ gauge coupling as
\begin{equation}
V_{\rm th2}
= a_{\rm th} \alpha_s^2 (T) T_{\rm p}^4 \ln ( | \phi_a |^2 / T_{\rm p}^2 )
\end{equation}
with $ a_{\rm th} \sim 1 $ depending on the particle contents.
The Hubble parameter $ H_{\rm th2} $
for the thermal log terms to be comparable to the Hubble induced mass terms
is also estimated with $ | \phi_a | \sim {\sqrt{H ( M / \lambda )}} $ as
\begin{equation}
H_{\rm th2}
\sim \alpha_s \left[ T_R^2 M_{\rm P} / (M / \lambda) \right]^{1/2} .
\end{equation}
Then, the thermal effects become important
for the Hubble parameter
\begin{equation}
H_{\rm th} \sim \max \left[ H_{\rm th1} , H_{\rm th2} \right] .
\end{equation}
It really takes the maximal value as
\begin{eqnarray}
H_{\rm th}^{\rm max} & \sim & H_{\rm th2} / \alpha_s
\nonumber \\
& \sim & 5 \times 10^8 {\rm GeV} \left( \frac{T_R}{10^9 {\rm GeV}} \right)
\left( \frac{m_{\nu_1}}{10^{-6} {\rm eV}} \right)^{1/2}
\label{Hth-max}
\end{eqnarray}
with certain value of the relevant coupling,
\begin{equation}
y \sim 3 \times 10^{-3} \left( \frac{T_R}{10^9 {\rm GeV}} \right)^{1/4}
\left( \frac{m_{\nu_1}}{10^{-6} {\rm eV}} \right)^{3/8} ,
\label{y}
\end{equation}
where Eq. (\ref{nmass}) is considered for $ m_{\nu_1} $ and $ M / \lambda $.
It is readily seen that $ H_{\rm osc} > H_{\rm th} $
for a wide range of the model parameters.
Then, the leptogenesis is completed dominantly
by the $ {\bar h}_e $ quartic term.
On the other hand, one may obtain
$ H_{\rm th} \gtrsim H_{\rm osc} $ for certain model parameter range
with $ {\bar h}_e \sim 10^{-5} $,
$ ( M / \lambda ) \lesssim 10^{18} {\rm GeV} $
and $ T_R \gtrsim 10^9 {\rm GeV} $.
Then, the oscillation of AD flaton fields is driven
by the thermal terms rather than the $ {\bar h}_e $ quartic term.
In any case, the lepton number asymmetry fluctuating after the inflation
is fixed to some significant non-zero value
in the early epoch with $ H \gg m_{3/2} $.
The thermal terms do not severely suppress the lepton number generation,
but even play a cooperative role in the flat manifold leptogenesis.

\section{Numerical Analysis}
\label{sec:numerical}

We here present the results of detailed numerical calculations
for the flat manifold leptogenesis.
The characteristic features in the multi-dimensional motion
of the AD-flaton fields are confirmed,
and the reasonable range of the model parameters
is identified for the sufficient leptogenesis.

The values of the various model parameters
are taken in the following range:
\begin{eqnarray}
&& M = 10^{17} {\rm GeV} , \
H_{\rm inf} = 10^{14} {\rm GeV} , \ t_0 = (2/3) H_{\rm inf}^{-1} ,
\nonumber \\
&& {\bar h}_e = 10^{-5} - 10^{-3} , \
T_R = 10^5 {\rm GeV} - 10^{10} {\rm GeV} ,
\nonumber \\
&& \lambda_{L \! \! \! /} , \lambda_H = 0.1 \lambda - 10 \lambda , \
\lambda = 10^{-3} - 1 ,
\nonumber \\
&& | a_{L \! \! \! /} | , | a_H | , c_a = 0 - 2 ,
\nonumber
\end{eqnarray}
and $ [ 0 , 2 \pi ] $ for the phases of coupling parameters.
Here, the parameters relevant to the low-energy part
$ V_{\rm low} $ are not presented explicitly.
It has been checked numerically that the effects of $ V_{\rm low} $
is negligible in the present scenario of flat manifold leptogenesis.
While the fixed value of the mass scale $ M $ in $ W_{\rm non} $
is presented for definiteness, the results are really obtained
as a function of $ M / \lambda $.
The value of $ \tan \beta $ is determined in Eq. (\ref{h_e})
for given $ {\bar h}_e $ with $ \theta_{12} \simeq 0 $
(small $ \nu_e $-$ \nu_\mu $ mixing)
or $ \theta_{12} \simeq \pi / 4 $ (large $ \nu_e $-$ \nu_\mu $ mixing).

By choosing randomly the model parameter values,
we have first determined the initial values of the AD-flaton fields
in Eq. (\ref{phit0}) just after the inflation.
(The initial phase of $ H_u $ is chosen
as $ \theta_{H_u}^{(0)} = 0 $ without loss of generality
by making a $ {\rm U(1)}_Y $ transformation.)
Then, we have solved the equations of motion (\ref{eqofm})
for a sufficiently long time interval
from $ t = t_0 $ ($ H = H_{\rm inf} $) to $ t \sim 10^8 t_0 $,
evaluating the particle number asymmetries $ \epsilon_a (t) $
as functions of time.
(In practice, we have solved Eq. (\ref{eqofmchi}) for $ \chi_a $
with Eq. (\ref{chi0}) as functions of $ z = \ln ( t / t_0 ) $,
since the time interval ranges over many orders.
The $ D $-flat condition (\ref{Dflat}) is checked to be hold
within numerical errors.)
The dominant contributions to the thermal terms
$ V_{\rm th1} + V_{\rm th2} $ are also taken into account
for the evolution of the AD-flaton fields.
The relevant parameters are taken for the thermal mass terms
as $ y = m_q / \langle H_d \rangle $ and $ c_{\rm th} = 3/4 $
with the $ q = d , s $ quarks,
$ y = m_c / \langle H_u \rangle $ and $ c_{\rm th} = 3/4 $
with the $ c $ quark,
and $ y = m_\mu / \langle H_d \rangle $ and $ c_{\rm th} = 1/4 $
with the muon.
These Yukawa couplings may be close to the optimal value
as given in Eq. (\ref{y}), depending on $ \tan \beta $.
As for the thermal log term, $ a_{\rm th} = 9/8 $ is taken
due to the decoupling of the top quark from the thermal plasma.

We have actually made calculations by taking hundreds of the parameter sets.
Some typical cases are listed in Table \ref{tab:typical},
where the sufficient lepton number asymmetry is obtained.
The cases (1) and (2) correspond
to the small $ {\bar h}_e $ (small $ \nu_e $-$ \nu_\mu $ mixing)
and the large $ {\bar h}_e $ (large $ \nu_e $-$ \nu_\mu $ mixing),
respectively.
Although different values are taken for $ M / \lambda $ and $ T_R $
in the cases (1) and (2), the resultant $ \epsilon_L $
has only moderate dependence on these parameters.
It should here be noticed that $ H_{\rm osc} \sim H_{\rm inf} $
in the case (2).
Then, the $ {\bar h}_e $ quartic term as well as the high-energy part
$ V_{\rm high} $ may be important to determine the potential minimum
in the inflation epoch.
We have really confirmed that even in such an extreme case
the multi-dimensional motion of the AD-fields can be realized
producing the significant lepton number asymmetry.

We have also searched the specific parameter range
allowing for somewhat high $ T_R \sim 10^{10} {\rm GeV} $,
where the sufficient lepton number asymmetry is obtained
with relatively large $ m_{\nu_1} \sim 10^{-4} {\rm eV} $
for the lightest neutrino.
A typical example is presented as the case (3) in Table \ref{tab:typical}.

\begin{figure}[ht]
\begin{center}
\includegraphics*[-3cm,0cm][10.0cm,22cm]{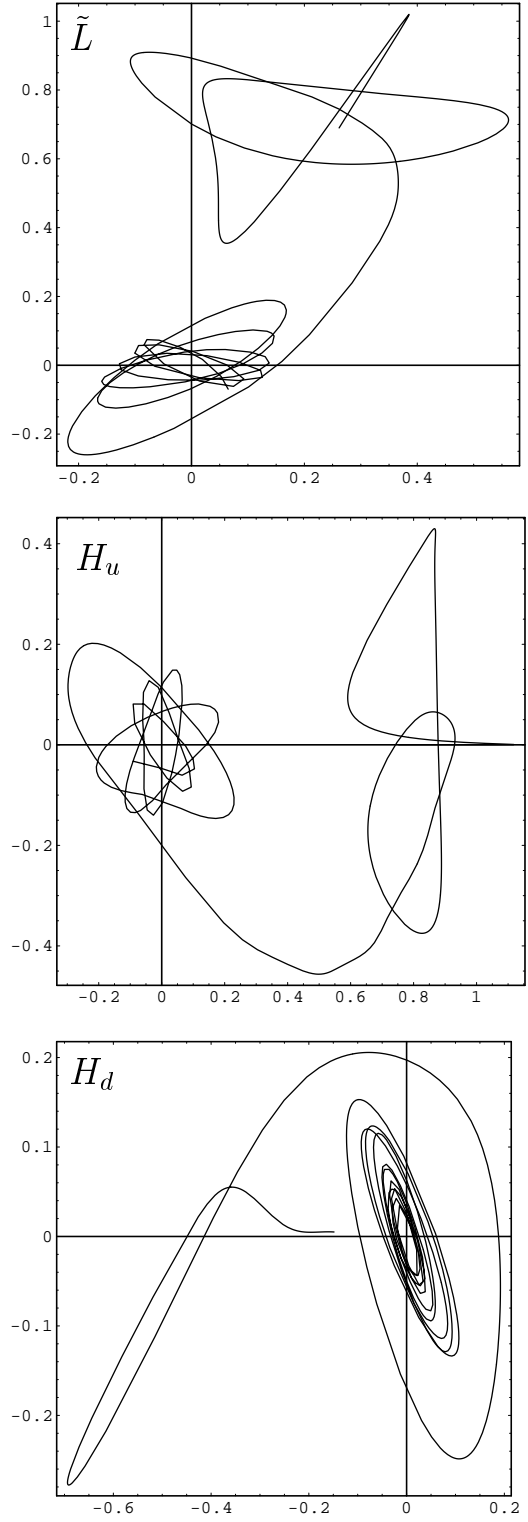}
\caption{The motions of the AD-flaton fields,
the real part (horizontal axis) and imaginary part (vertical axis),
are depicted in terms of the dimensionless fields $ \chi_a $
for the case (1) in Table \ref{tab:typical}.
}
\label{trace}
\end{center}
\end{figure}

\begin{figure}[ht]
\begin{center}
\includegraphics*[0cm,0cm][24.0cm,15.5cm]{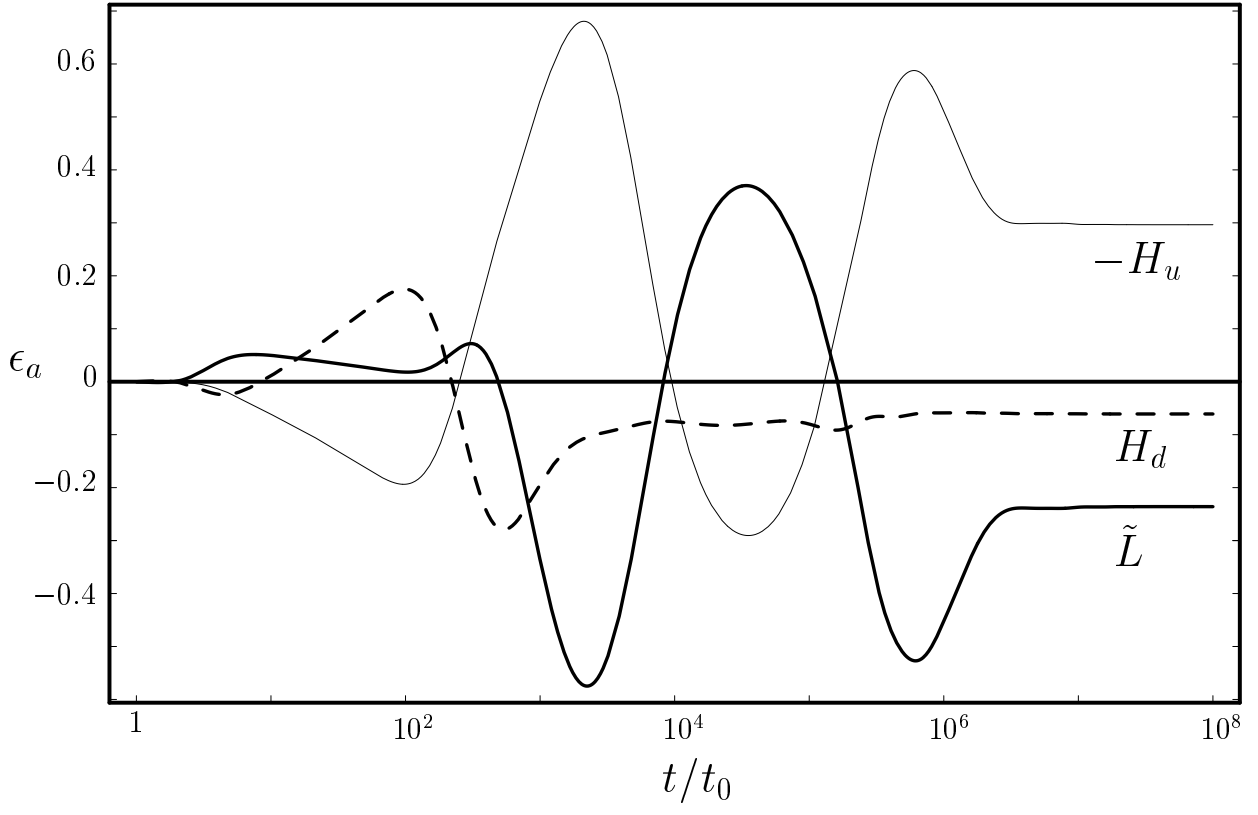}
\vspace*{0.5cm}
\caption{The time variation of particle number asymmetries
is shown for the case (1) in Table \ref{tab:typical}.
}
\label{epsilon}
\end{center}
\end{figure}

The time evolution of the AD-flaton fields has been determined
precisely by these numerical calculations.
In Fig. \ref{trace}, the multi-dimensional motion of the AD-flaton fields
is typically depicted in terms of the dimensionless fields
$ \chi_a $ for the case (1) in Table \ref{tab:typical}.
In Fig. \ref{epsilon}, is also shown the time variation
of the particle number asymmetries,
or the contributions to the hypercharge asymmetry,
$ \epsilon_{\tilde L} $ (bold line),
$ - \epsilon_{H_u} $ (slim line)
and $ \epsilon_{H_d} $ (dashed line),
which are evaluated with the solutions of $ \chi_a $.
Here, we can check the hypercharge conservation
(within the numerical errors),
$ \epsilon_{\tilde L} - \epsilon_{H_u} + \epsilon_{H_d} = 0 $.
The lepton number asymmetry is given just
by $ \epsilon_L = \epsilon_{\tilde L} $.
These asymmetries are really fluctuating after the inflation,
and then fixed to certain values for $ t \gg H_{\rm osc}^{-1} $.

\begin{figure}[ht]
\begin{center}
\includegraphics*[0cm,0cm][24.0cm,15.5cm]{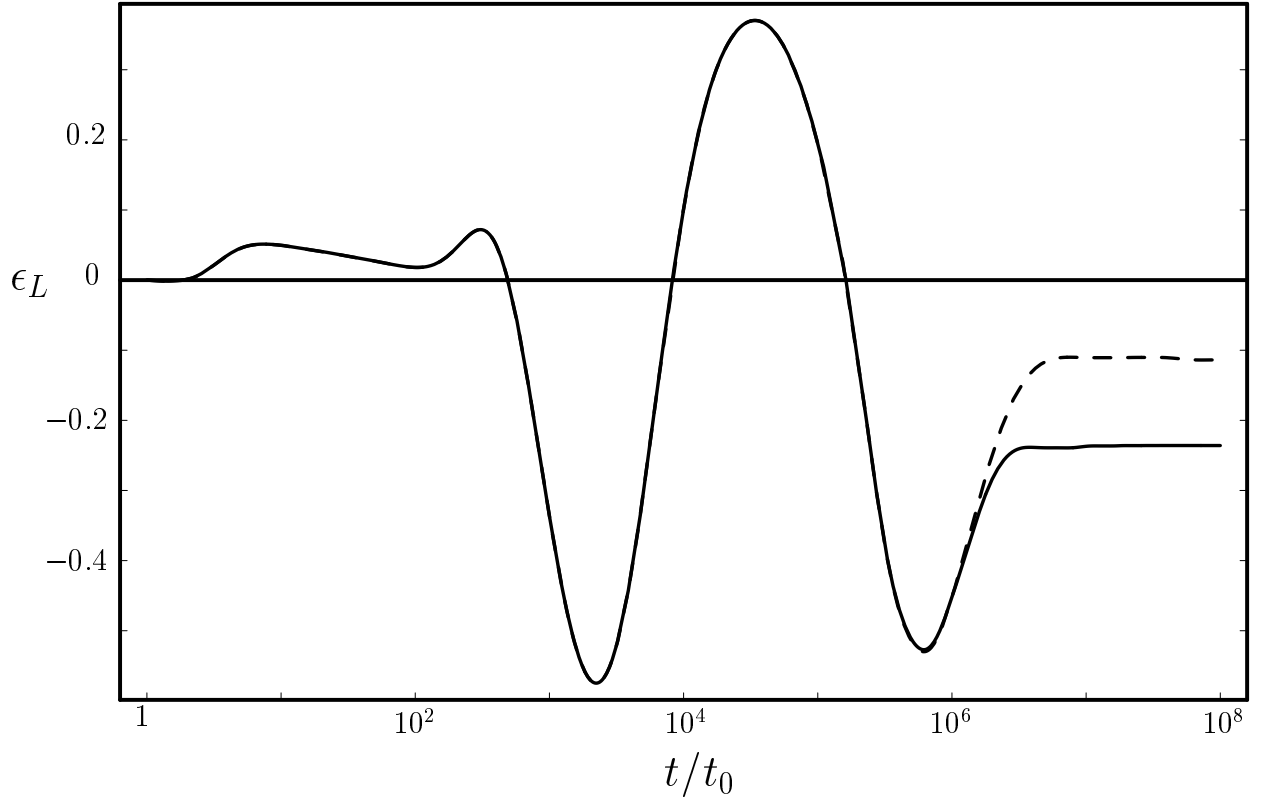}
\vspace*{0.5cm}
\caption{
The time variation of the lepton number asymmetry is shown
for the case (1) in Table \ref{tab:typical},
where the solid and dashed lines represent the results
with and without the thermal terms, respectively.
}
\label{asym-th}
\end{center}
\end{figure}

As for the thermal effects, we have observed
that they do not alter essentially
the present scenario of flat manifold leptogenesis.
In a wide range of the model parameter space,
the thermal effects on the final lepton number asymmetry $ \epsilon_L $
are found to be small for $ H_{\rm osc} \gg H_{\rm th} $,
particularly with lower reheating temperature
$ T_R \lesssim 10^8 {\rm GeV} $.
It should, however, be noted
that the lepton number asymmetry varies slowly
with $ z = \ln ( t / t_0 ) $, as seen in Fig. \ref{epsilon}.
Then, it actually takes a rather long term ranging over some orders
around $ t \sim H_{\rm osc}^{-1} $
to fix completely the lepton number asymmetry
by driving the AD-flaton oscillation.
In such a situation, 
when $ H_{\rm th}^{\rm max} $ is smaller only by a few orders
than $ H_{\rm osc} $, as seen for the case (1) of Table \ref{tab:typical},
the thermal terms become dominant at the late stage of leptogenesis
for driving the AD-flaton oscillation.
In Fig. \ref{asym-th}, the time variation of the lepton number asymmetry
$ \epsilon_L (t) $ is shown for the case (1) in Table \ref{tab:typical},
where the solid and dashed lines represent
the results with and without the thermal terms, respectively.
In this case, the thermal log term is considered
to provide the dominant effect.
We really observe here that the lepton number asymmetry is
finally fixed by the thermal terms,
though the $ {\bar h}_e $ quartic term first triggers
the AD-flaton oscillation.
The resultant $ \epsilon_L $ is changed by some factor $ \sim 1 $
due to the thermal effects.
We may even have $ H_{\rm osc} \lesssim H_{\rm th}^{\rm max} $
for some cases with smaller $ {\bar h}_e \sim 10^{-5} $,
$ M / \lambda \lesssim 10^{18} {\rm GeV} $
and higher $ T_R \gtrsim 10^9 {\rm GeV} $.
Then, the AD-flaton oscillation to complete the leptogenesis
is driven mainly by the thermal terms
rather than the $ {\bar h}_e $ quartic term.
In any case, the thermal terms do not provide severe suppression,
but may even play a cooperative role for the flat manifold leptogenesis.

\begin{figure}[ht]
\begin{center}
\includegraphics*[0cm,0cm][24.0cm,15.5cm]{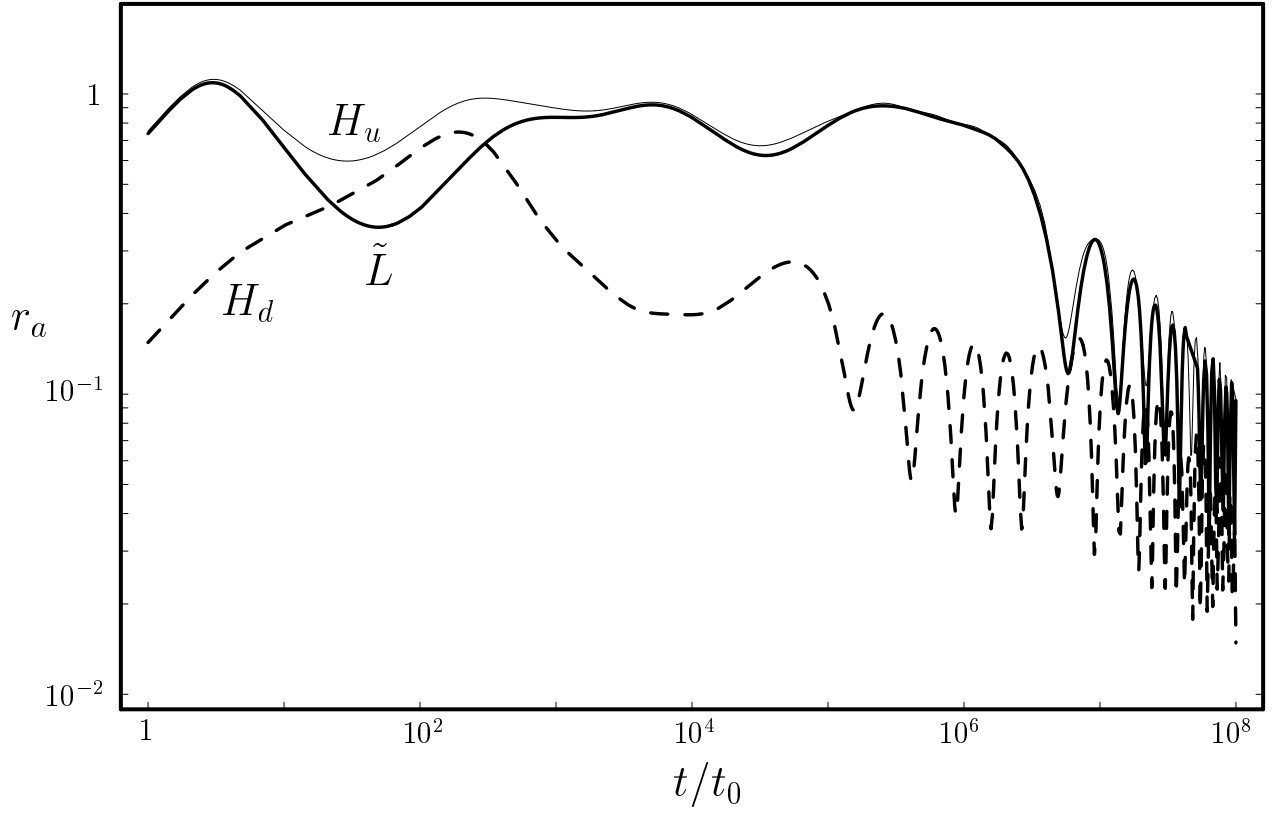}
\vspace*{0.5cm}
\caption{
The time variation of the AD-flaton field magnitudes
is shown in terms of $ r_a $
for the case (1) in Table \ref{tab:typical}.
}
\label{chi}
\end{center}
\end{figure}

The magnitudes of AD-flaton fields are found
to be scaled roughly as
\begin{equation}
| \phi_a | \propto H^{\alpha (t)} .
\end{equation}
In Fig. \ref{chi}, this redshift is shown typically
for the case (1) of Table \ref{tab:typical}
in terms of the dimensionless variables
$ r_a (t)  \propto H^{\alpha (t) - 1/2} $.
It is observed up to some fluctuating behavior
that the power of redshift $ \alpha (t) $ eventually changes as
\begin{equation}
\alpha (t) \approx
\left\{ \begin{array}{cl}
1/2 & ( t < H_{\rm osc}^{-1} ) \\
2/3 & ( H_{\rm osc}^{-1} \lesssim t < H_{\rm th}^{-1} ) \\
7/8 & ( H_{\rm th}^{-1} \lesssim t < m_{3/2}^{-1} ) \\
1 & ( t \gtrsim m_{3/2}^{-1} ) \end{array} \right. .
\label{alpchange}
\end{equation}
We here find especially
that in the late epoch $ H_{\rm th}^{-1} \lesssim t < m_{3/2}^{-1} $
the evolution of the AD-flatons is determined
dominantly by the thermal terms,
while the leptogenesis is already completed
during the epoch $ H_{\rm osc}^{-1} \lesssim t \lesssim H_{\rm th}^{-1} $.
The redshift of the AD-flaton fields changes finally
to $ | \phi_a | \propto H $,
when the low-energy soft supersymmetry breaking mass terms
become dominating.

\begin{figure}[ht]
\begin{center}
\includegraphics*[0cm,0cm][24.0cm,15.5cm]{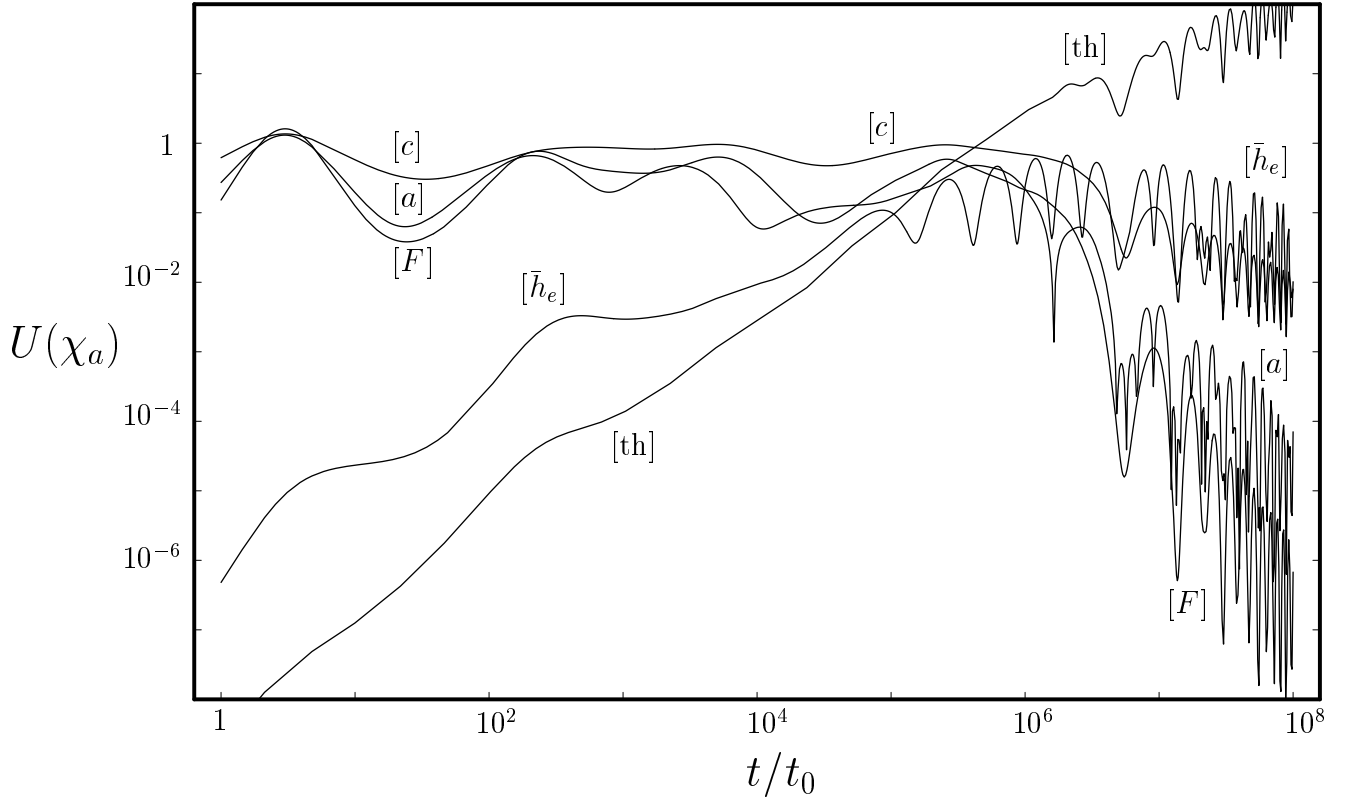}
\vspace*{0.5cm}
\caption{
The time variation of the scalar potential terms
is shown in terms of $ U( \chi ) $
for the case (1) in Table \ref{tab:typical}.
The respective terms are indicated as
$ [ c ] $: Hubble induced negative mass-squared terms,
$ [ a ] $: Hubble induced $ A $ terms,
$ [ F ] $: $ | F |^2 $ terms from $ W_{\rm non} $,
$ [ {\bar h_e} ] $: $ | {\tilde L} H_d |^2 $ term,
$ [ {\rm th} ] $: thermal terms.
}
\label{terms}
\end{center}
\end{figure}

In Fig. \ref{terms}, the time variation of the scalar potential terms
is also shown in terms of the dimensionless effective potential
$ U ( \chi_a ) $ in Eq.(\ref{U}),
where the symbols indicate the respective terms as
$ [ c ] $: Hubble induced negative mass-squared terms,
$ [ a ] $: Hubble induced $ A $ terms,
$ [ F ] $: $ | F |^2 $ terms from $ W_{\rm non} $,
$ [ {\bar h_e} ] $: $ | {\tilde L} H_d |^2 $ term,
$ [ {\rm th} ] $: thermal terms.
It is observed that in the epoch with $ H > H_{\rm osc} $
the $ [ c ] $, $ [ a ] $ and $ [ F ] $ terms
in $ V_{\rm high} $ really dominate being scaled as $ H^0 $
in terms of the $ U ( \chi_a ) $.
Then, the $ [ {\bar h_e} ] $ and even $ [ {\rm th} ] $ terms
catches up them around $ H \sim H_{\rm osc} $.
Soon after that, the $ [ {\rm th} ] $ term dominates over
the $ [ {\bar h}_e ] $ and $ [ c ] $ terms,
and the $ [ a ] $ and $ [ F ] $ terms
including the lepton number violation decrease rapidly,
so that the lepton number asymmetry is fixed to some non-zero value.
The $ D^2 $ term is not shown here for simplicity.
Since $ D = ( g^2 + g^{\prime 2} )^{1/2}
( | {\tilde L} |^2 - | H_u |^2 + | H_d |^2 ) $
is calculated by using the solutions of $ \phi_a (t) $,
it is very sensitive to the numerical errors
for the cancellation among $ | \phi_a |^2 $ terms by several orders.
We have really checked that the $ D^2 $ contribution
is smaller than $ 10^{-3} $ in the unit of $ U( \chi_a ) $,
though it apparently exhibits a violent oscillation
within this small range.
This oscillating behavior is regarded to be an artifact
within the intrinsic numerical errors
due to the fine cancellation among the large $ | \phi_a |^2 $ terms
in calculating the $ D $ term.
The $ D^2 $ term is anyway small enough
compared to the leading terms in the $ U( \chi_a ) $,
and the $ D $-flat condition (\ref{Dflat}) is maintained
quite well through the evolution of the AD-flaton fields.

It has been argued that the comparability
of $ \lambda_{L \! \! \! /} $ and $ \lambda_H $
is essential for the flat manifold leptogenesis.
In Fig. \ref{scatter}, the resultant lepton number asymmetries are plotted
depending on the ratio $ \lambda_{L \! \! \! /} / \lambda_H $,
where the relevant parameter values are taken randomly.
It is clearly seen that
the flat manifold leptogenesis can be realized naturally
under the flatness condition (\ref{flatcond}).

\section{Conclusion}
\label{sec:conclusion}

We have investigated the flat manifold leptogenesis a la Affleck-Dine
with the slepton and Higgs fields, $ \tilde{L} $, $ H_u $, $ H_d $,
in the supersymmetric standard model.
The multi-dimensional motion of these AD-flaton fields is indeed realized
in the case that the $ \tilde{L} H_u $ and $ H_u H_d $ directions
are comparably flat with the relevant non-renormalizable
superpotential terms.
Soon after the inflation, the lepton number asymmetry
appears to fluctuate due to this multi-dimensional motion
involving certain $ CP $ violating phases.
Then, the lepton number asymmetry is fixed
to some significant non-zero value for the successful baryogenesis
when the scalar fields begin to oscillate with rotating phases
driven by the quartic coupling from the superpotential term
$ {\bar h}_e L H_d e^c $ with $ {\bar h}_e \sim 10^{-5} - 10^{-3} $.
The Hubble parameter  $ H_{\rm osc} $ at this epoch
for the completion of leptogenesis is much larger
than the gravitino mass $ m_{3/2} \sim 10^3 {\rm GeV} $.
The thermal terms do not alter this scenario of flat manifold leptogenesis
in the early epoch.
They may even play a cooperative role for leptogenesis.
The lightest neutrino mass can be $ m_{\nu_1} \sim 10^{-4} {\rm eV} $,
if the reheating temperature is allowed to be
$ T_R \sim 10^{10} {\rm GeV} $.
Clearly, this flat manifold leptogenesis is not restricted
by the physics at the electroweak scale
such as the low-energy supersymmetry breaking terms.

\begin{acknowledgments}
This work is supported in part by
Grant-in-Aid for Scientific Research on Priority Areas B (No. 13135214)
from the Ministry of Education, Culture, Sports, Science and Technology,
Japan.
\end{acknowledgments}

\begin{table*}
\caption{
\label{tab:typical}
Typical cases for the flat manifold leptogenesis,
where $ M = 10^{17} {\rm GeV} $ and $ H_{\rm inf} = 10^{14} {\rm GeV} $
are taken for definiteness.
The index $ a $ denotes the AD-flatons
as $ \phi_a = {\tilde L} , H_u , H_d $ in order.
}
\begin{ruledtabular}
\begin{tabular}{cccc}
 & case (1) & case (2) & case (3)
\\
\hline
\\
$ {\bar h}_e $
 & $ 3 \times 10^{-5} $
 & $ 10^{-3} $
 & $ 10^{-3} $
\\
$ ( \tan \beta , \theta_{12} ) $
 & $ ( 10.2 , 0 ) $
 & $ ( 2.1 , \pi / 4 ) $
 & $ ( 2.1 , \pi / 4 ) $
\\
$ ( M / \lambda , T_R ) $
 & $ ( 10^{19} {\rm GeV} , 10^9 {\rm GeV} ) $
 & $ ( 10^{20} {\rm GeV} , 10^8 {\rm GeV} ) $
 & $ ( 10^{18} {\rm GeV} , 10^{10} {\rm GeV} ) $
\\
$ ( H_{\rm osc} , H_{\rm th}^{\rm max} ) $
 & $ ( 9 \times 10^9 {\rm GeV} , 5 \times 10^8 {\rm GeV} ) $
 & $ ( 10^{14} {\rm GeV} , 10^7 {\rm GeV} ) $
 & $ ( 10^{12} {\rm GeV} , 10^{10} {\rm GeV} ) $
\\
$ ( \lambda_{L \! \! \! /} , \lambda_H ) $
 & $ ( 1.0 \times 10^{-2} , 1.5 \times 10^{-2} ) $
 & $ ( 1.5 \times 10^{-3} , 0.5 \times 10^{-3} ) $
 & $ ( 4.0 \times 10^{-1} , 1.0 \times 10^{-1} ) $
\\
$ ( a_{L \! \! \! /} , a_H ) $
 & $ ( 1.0 {\rm e}^{i (1/4) \pi} ,
       1.0 {\rm e}^{i (1/2) \pi} ) $
 & $ ( 1.5 {\rm e}^{-i (1/3) \pi} ,
       0.5 {\rm e}^{-i \pi} ) $
 & $ ( 2.0 {\rm e}^{i (1/3) \pi} ,
       0.4 {\rm e}^{i \pi} ) $
\\
$ c_a $
 & $ ( 1.5 , 1.0 , 0.5 ) $
 & $ ( 1.0 , 1.0 , 1.0 ) $
 & $ ( 2.0 , 2.0 , 0.5 ) $
\\
$ {\displaystyle{ \left( \begin{array}{c}
r_a^{(0)} \\ \theta_a^{(0)} \end{array} \right) }} $
 & $ {\displaystyle{ \left( \begin{array}{rrr}
      0.738 & 0.753 & 0.148 \\
      1.208 & 0 & 3.108 \end{array} \right) }} $
 & $ {\displaystyle{ \left( \begin{array}{rrr}
      0.397 & 0.954 & 0.868 \\
      -1.170 & 0 & 0.237 \end{array} \right) }} $
 & $ {\displaystyle{ \left( \begin{array}{rrr}
      0.313 & 0.640 & 0.558 \\
      -2.040 & 0 & 2.783 \end{array} \right) }} $
\\
$ {\displaystyle{ \left( \begin{array}{c}
r_a^{(1)} \\ \theta_a^{(1)} \end{array} \right) }} $
 & $ {\displaystyle{ \left( \begin{array}{rrr}
      0.775 & 0.823 & 0.277 \\
      1.236 & 0.039 & 3.101 \end{array} \right) }} $
 & $ {\displaystyle{ \left( \begin{array}{rrr}
      0.392 & 1.047 & 0.971 \\
      -1.188 & -0.007 & 0.226 \end{array} \right) }} $
 & $ {\displaystyle{ \left( \begin{array}{rrr}
      0.282 & 0.711 & 0.654 \\
      -2.019 & 0.005 & 2.819 \end{array} \right) }} $
\\
$ m_{\nu_1} $
 & $ 3.0 \times 10^{-6} {\rm eV} $
 & $ 3.7 \times 10^{-7} {\rm eV} $
 & $ 1.0 \times 10^{-4} {\rm eV} $
\\
$ ( \epsilon_L , n_L / s ) $
 & $ ( -0.24 , -2.1 \times 10^{-10} ) $
 & $ ( -0.17 , -1.5 \times 10^{-10} ) $
 & $ ( -0.11 , -1.0 \times 10^{-10} ) $
\\
\end{tabular}
\end{ruledtabular}
\end{table*}

\clearpage

\newpage

\newpage

\newpage

\newpage

\newpage

\begin{figure}[ht]
\begin{center}
\hspace*{0.5cm}
\includegraphics*[0cm,0cm][24.0cm,15.5cm]{scatter.ps}
\vspace*{0.5cm}
\caption{
A scatter plot is presented
for the resultant lepton number asymmetries
depending on the ratio $ \lambda_{L \! \! \! /} / \lambda_H $,
where the relevant parameter values are taken randomly.
}
\label{scatter}
\end{center}
\end{figure}


\begin{thebibliography}{99}

\bibitem{AD}
I. Affleck and M. Dine,
Nucl. Phys. {\bf B 249}, 361 (1985).

\bibitem{DRT}
M. Dine, L. Randall and S. Thomas,
Nucl. Phys. {\bf B 458}, 291 (1996).

\bibitem{LHu}
H. Murayama and T. Yanagida,
Phys. Lett. {\bf B 322}, 349 (1994);
T. Moroi and H. Murayama,
JHEP {\bf 07}, 009 (2000);
M. Fujii, K. Hamaguchi and T. Yanagida,
Phys. Rev. {\bf D 65}, 043511 (2002). 

\bibitem{LHu2}
T. Asaka, M. Fujii, K. Hamaguchi and T. Yanagida,
Phys. Rev. {\bf D 62}, 123514 (2000).

\bibitem{LHu3}
M. Fujii, K. Hamaguchi and T. Yanagida,
Phys. Rev. {\bf D 63}, 123513 (2001);
hep-ph/0203189.

\bibitem{myletter}
M. Senami and K. Yamamoto,
Phys. Lett. B. {\bf 524}, 332 (2002).

\bibitem{thermaleffect}
R. Allahverdi, B. A. Campbell and J. Ellis,
Nucl. Phys. {\bf B579}, 355 (2000);
A. Anisimov and M. Dine,
Nucl. Phys. B {\bf 619}, 729 (2001).

\bibitem{seesaw}
T. Yanagida, in Proceedings of {\it Workshop on Unified Theory
and Baryon number in the universe}, eds.
O. Sawada and A. Sugamoto (KEK 1979);
M. Gell-Mann, P. Ramond and R. Slansky,
in {\it Supergravity}, eds. P. van Niewenhuizen
and D. Z. Freedman (North Holland 1979).

\bibitem{flaton}
K. Yamamoto,
Phys. Lett. {\bf B 161}, 289 (1985);
{\it ibid.} {\bf B 168}, 341 (1986).

\bibitem{harveyturner}
J. A. Harvey and M. S.Turner, Phys. Rev. {\bf D 42}, 3344 (1990).

\bibitem{eta}
D. E. Groom {\it et al.}, Particle Data Group,
Eur. Phys. J. {\bf C 15}, 1 (2000), http://pdg.lbl.gov/.

\bibitem{gravitino}
J. Ellis, J. E. Kim and D. V. Nanopolous,
Phys. Lett. {\bf B 145}, 181 (1984);
E. Holtmann, M. Kawasaki, K. Kohri and T. Moroi,
Phys. Rev. {\bf D 60}, 023506 (1999);
M. Kawasaki, K. Kohri and T. Moroi,
{\it ibid.} {\bf D 63}, 103502 (2001).

\bibitem{gravitino2}
K. Kohri,
Phys. Rev. {\bf D 64}, 043515 (2001).

\bibitem{gravitino3}
R. Kallosh, L. Kofman, A. Linde and A. V. Proeyen,
Phys. Rev. {\bf D 61}, 103503 (2000);
G.F. Giudice, I. Tkachev and A. Riotto,
JHEP {\bf 08}, 009 (1999); {\it ibid} {\bf 11}, 036 (1999);
H. P. Nilles, M. Peloso and L. Sorbo,
Phys. Rev. Lett {\bf 87}, 051302 (2001).

\end{thebibliography}
\end{document}